\documentstyle[prl,preprint,aps]{revtex}

\begin{document}
\draft
\title{The overlapping of nonlinear resonances and the problem of quantum chaos}
\author{A. Ugulava, L. Chotorlishvili, K. Nickoladze.}
\address{Tbilisi State University,Department of Physics,\\
Chavchavadze av. 3, 0128 Tbilisi, Georgia}
\date{\today}
\maketitle

\begin{abstract}
The motion of nonlinearly oscillating particle under the influence of a
periodic sequence of short impulses has been investigated. Analysis of the
Schrodinger equation for the universal Hamiltonian has been done. It is
shown that the quantum criterion of overlapping of resonances is of form $%
\lambda K\geq 1$, where $K$-is the classical coefficient of stochasticity ,
and $\lambda $ is the functional defined with the using of Mathieu
functions.The area of maximal values of $\lambda \ $is determined. The idea
about emerging of quantum chaos originated due to the adiabatic motion along
the curves of Mathieu-characteristics at multiple passages through the
points of branching is advanced.
\end{abstract}





\section{Introduction}

The overlapping of nonlinear resonances is the criterion for the origin of
dynamical stochasticity in classical Hamiltonian systems. At realization of
the condition of this criterion it is possible to justify the transition
from the dynamic Hamilton description to the statistical one and to study
the behavior of the system with the help of statistical average. Such
description is as full as possible in this range and successfully
substitutes dynamic description which loses its sense due to strong local
instability \cite{1,2}. However in the quantum mechanics the introduction of
a stochasticity is significantly difficult \cite{3,4,5,6,7,8}. What can be
considered as quantum analog of dynamic stochasticity, what is a criterion
for passing to quantum chaos, how to quantize the system in classical limit
corresponding to the dynamic stochasticity? There are only some problems of
quantum chaos.

In the given work the attempt is made to investigate two aspects of a
general problem of quantum chaos: the criterion for the overlapping of
resonances on the basis of quantum mechanics and to study the singularities
of wave functions in the area, in which the classic mechanics assumes the
existence of dynamic stochasticity.

Let us assume, that nonlinearly oscillating particle (Fig. 1) is under the
action of variable field 
\begin{equation}
H(x,p)=H_{o}(x,p)+H_{NL}(x)+\varepsilon V(x,t), \\
\end{equation}
where 
\[
H_o(x)=1/2(p^2/m+\omega_o^2 mx^2),
\]
\[
H_{NL}=\gamma x^3+\beta x^4+\ldots,
\]
\[
\varepsilon V(x,t)=-(e/m) xf(t),
\]
\[
f(t)=f_o\Sigma(t)\cos{\omega t},\;\;\; \varepsilon V(x,t)=\varepsilon
V_ox\Sigma(t)\cos{\omega t},
\]
\[
\varepsilon V_o=-(e/m)f_o,
\]
\begin{equation}
\varepsilon\ll1. \label{2}
\end{equation}

The Hamiltonian of such type for a long time is being investigated with the
purpose of study of a dynamic stochasticity both in classical \cite{1}, and
in quantum systems \cite{3,5,6}.

Here $x$ and $p$ are the coordinate and the impulse of the particle, $%
\omega_o$ is the fundamental frequency,$\gamma$ and $\beta$ are the
coefficients of the nonlinearity,$m$ and $e$ are the mass and the charge of
the particle,$f_o$ is the amplitude of the variable field, $\Sigma(t)$ is
the periodic sequence of rectangular electromagnetic impulses with the
duration $\tau$ and with the phase of recurring $T$ (Fig.2,Fig.3). It is
supposed, that $1/\omega,1/\omega_o\ll\tau\ll T.$

The fundamental component of pumping field at frequency $\omega=\omega_o$ is
able to carry out the linear resonance and cause the increase of $x$ until
the nonlinear terms proportional to $x^3$ and $x^4$ become significant in
the potential (i.e. up to a neighborhood of $x_L$, Fig. 1). From this moment
the nonlinear terms will gradually begin to detune the linear resonance (at $%
\omega=\omega_o$), that will reduce the resonance growth of $x$. Then the
remaining harmonics of the pumping spectrum, concentrated in sum $\Sigma(t)$%
, will begin to play the role. Their role will be significant in reaching
higher excitation $(x>x_L)$, if criteria of the overlapping of resonances is
fulfilled.

\section{Universal Hamiltonian. Classical consideration}

In this paragraph we shall review the well-known results obtained in the
theory of a stochasticity for the nonlinearly oscillating classical systems.
After passing in the Hamiltonian (1) to the variables action - angle with
help of transformation $x=\sqrt{2I/m\omega _{o}}\cos {\theta },p=-\sqrt{%
2I\omega _{o}m}\sin {\theta }$, and averaging (1) with respect to the fast
phase $\theta $, we obtain: 
\[
H=H_o^{NL}+\varepsilon V(I)\ldots ,
\]
\[
H_o^{NL}=H_o+H_{NL},
\]
\[
H_o=I\omega_o, H_{NL}=3\pi\beta(I/m\omega_o)^2,
\]
\begin{equation}
\varepsilon V(x,t)\!=\!\varepsilon/2 V(I)\cos{\varphi} \Sigma(t)\equiv%
\varepsilon V(I,\varphi,t),\;V(I)\!=\!V_o \sqrt{2I/m\omega_o}  \label{3}
\end{equation}
where $\varphi =\theta -\omega t$ is the slow phase.

Let us notice, that $H_{NL}$ we have united with $H_o$ in unperturbed
Hamiltonian $H_o^{NL}$. In the further nonlinear terms are not assumed small
and the application for them of the perturbation theory is not possible. The
relevant set of canonical equation looks like: 
\[
{\dot I}=-\varepsilon\frac{\partial V(I,\varphi,t)}{\partial\phi},
\]
\[
\dot\varphi=\omega(I)+\varepsilon\partial V(I,\varphi,t)/\partial I,
\]
where 
\begin{equation}
\omega(I)=\omega_o-\omega+\omega_{{}_{NL}}(I), \omega_{{}_{NL}}= 6\pi \beta
I/ (m^2 \omega_o^2),  \label{4}
\end{equation}
\begin{equation}
\varepsilon V(I,\varphi,t)=1/2\frac{\tau}{T}V(I)\cos\varphi
\sum_{-1/\tau}^{1/\tau}\cos k\Omega t.  \label{5}
\end{equation}

The phase $\varphi$, slow as compared to the $\theta$ remains fast in the
comparison with the velocity of the action $I$ variation. The velocity of
the variation of $\varphi$ contains information about the nonlinear
character of motion. In particular, the dependence of $\dot\varphi$ from $I$
means the presence of nonlinearity in the oscillating system. Suppose that
for some values of the $I_n$ the resonance between $\omega(I)$ and some
component from the polychromatic pumping spectrum (5) (i.e. $%
\omega(I_n)\approx n\Omega$) is carried out. Then forming a slow phase $%
\alpha_n\approx\varphi-n\Omega t$, averaging expression (5) with respect to
the fast $\varphi$ and taking into account (5), we get: 
\begin{equation}
\varepsilon V(I,\varphi,t)=1/4 \frac{\tau}{T}V(I)\cos\alpha_n.  \label{6}
\end{equation}

Substituting (6) in (4), we have: 
\begin{equation}  \label{7}
\dot{I_n}=U(I)\sin\alpha_n, \\
\end{equation}
\[
\dot{\alpha_n}=\omega(I_n)-n\Omega+dU(I)/dI \cos\alpha_n,
\]
where 
\begin{equation}
U(I)=1/4\frac{\tau}{T}V(I).  \label{8}
\end{equation}

Equations (7) describe the nonlinear resonance. As opposite to the linear
resonance at which unbounded linear growth of an amplitude is valid (in our
case action $I$ or deviation $x$), in the case of the nonlinear resonance
(as it was already mentioned), there are so-called "phase oscillations",
i.e. oscillation of the phase $\alpha_n$ and the amplitude $I_n$.

Let us introduce the deviation of the action, $\Delta I_n=I-I_n$, $\Delta
I\ll I_n$ from the resonance value. Then it is possible to demonstrate that
the Hamiltonian 
\begin{equation}
\tilde H= \omega^\prime(\Delta I)^2/2+U(I_n)\cos\alpha_n,  \label{9}
\end{equation}
where $\omega^\prime$ = $\Big(d\omega/dI\Big)_{I=I_n}$, produces the set of
equations of (7) type. Really, from the equilibrium conditions, $\dot\alpha=%
\dot I=0$, one can obtain:

\begin{equation}  \label{10}
\omega(I_n)-n\Omega+\varepsilon/2dU(I_n)/dI=0, \\
\end{equation}
\[
\omega(I_n)=\omega(I_n)+\omega^\prime\Delta I_n.
\]
If the condition of moderate nonlinearity is just $\varepsilon\ll\mu\ll1/%
\varepsilon$, where 
\begin{equation}
\mu=I_n/\omega(I_n)\biggl(\frac{d\omega}{dI}\biggr)_{I=I_n},  \label{11}
\end{equation}
is the factor of nonlinearity, then with the help of (7)-(11), we get 
\begin{equation}  \label{12}
\Delta\dot{I_n} =U(I_n)\sin\alpha_n, \\
\end{equation}

It is possible to obtain the equation $\dot{\alpha_n}=\omega^\prime\Delta I_n
$ for phase oscillations from the set (12) 
\begin{equation}
\ddot\alpha_n-\omega_{ph}^2 \sin\alpha_n=0,  \label{13}
\end{equation}
where $\omega_{ph}=\sqrt{\omega^\prime U}$ is the frequency of phase
oscillations.

Let us notice, that the Hamiltonian (9) is the Hamiltonian of the
mathematical pendulum, where the role of the mass plays $1/\omega^\prime$
and the role of the potential energy - $U(I_n)\cos\alpha_n$. Taking into
account that in the classical mechanics the problem of the pendulum can be
solved exactly we reduced an initial problem to the solved one.

Variation of the action with the help of (9) and (12) can be presented in
the form: 
\begin{equation}
\Delta I_{+}=\sqrt{(E+U)/\omega ^{\prime }}dn(\omega ^{\prime }\sqrt{%
(E+U)/\omega ^{\prime }}t;k),\;\;\;E>U  \label{14}
\end{equation}
(with the period equal to $2K(k)$), 
\begin{equation}
\hskip-1cm\Delta I_{-}=\sqrt{(E+U)/\omega ^{\prime }}cn(\omega ^{\prime }%
\sqrt{(E+U)/\omega ^{\prime }}t;1/k),\;\;\;E<U,  \label{15}
\end{equation}
(with the period equal to $4K(1/k)$), where $E$ is the energy of the
particle, $cn$ and $dn$ are Jacobian elliptic functions: the elliptic cosine
and delta of amplitude; $K(k)$ is the second order complete elliptic
integral, $k=\sqrt{2U/(E+U)}$ is the module of elliptic integrals. $\Delta
I_{+}$ and $\Delta I_{-}$ are deviations of the action up and below the
separatrix accordingly. For $E=U(\text{or}\;\;\;k\rightarrow 1)$ these two
solutions are sewed together and take the form of an instanton 
\begin{equation}
\Delta I_{+}\rightarrow \Delta I_{-}\rightarrow \frac{\sqrt{2U/\omega
^\prime }}{c\hbar(\sqrt{2U\omega ^{\prime }}t)}.  \label{16}
\end{equation}

Averaging the action deviation with respect to half period we get the
following equations: 
\[
\overline{\Delta I_+}=\sqrt{(E+U)/\omega^\prime}\frac{1}{K(k)}\int
\limits_{0}^{4K(k)}dn(\tau,k)d\tau=
\]
\begin{equation}  \label{17}
=\frac{\pi\sqrt{E+U/\omega^\prime}}{2K(k)},\;\;\;E>U; \\
\end{equation}
\[
\overline{\Delta I_-}=\sqrt{(E+U)/\omega^\prime}\frac{1}{2K(1/k)}\int
\limits_{\alpha_o}^{2K(1/k)}cn(k\tau,1/k)d\tau= 
\]
\begin{equation}
=\frac{\pi\sqrt{U/2\omega^\prime}}{K(1/k)},\;\;\;E<U;  \label{18}
\end{equation}
$\alpha_o=\arccos (-E/U)$.

At $E\approx U$ 
\begin{equation}
\overline{\Delta I_{-}}\approx \overline{\Delta I_{+}} \approx \overline{%
\Delta I}=\pi \sqrt{U/2}\frac{1}{\ln 4\sqrt{4U/(U-E)}}.  \label{19}
\end{equation}

Action variation dependence on the ratio $E/U$ is presented on the Fig.4.

According to Fig.4, the magnitude of $\Delta I_{-}$ sharply decreases with
approaching to a separatrix.

If during the phase oscillation $\Delta I_{n}$ takes enough major values
(such as $\omega ^{\prime }\Delta I_{n}\geq \Omega $), the resonance
condition $\omega (I_{n})\approx n\Omega $ breakes, but other resonance
condition is attuned: 
\begin{equation}
\omega (I_{n}^{o}+\Delta I_{n})\approx (n+1)\Omega .  \label{20}
\end{equation}

Just at the jump to the other resonance condition there is an abruptness,
which results in a stochastic wandering of spectral harmonics (6). It is the
essence of overlapping of resonances, which serves as the criterion of the
stochasticity emerging in the nonlinearly oscillating system. Expending $%
\omega (I)$ into series with respect to $\Delta I,$ and making an
estimation, $\Delta I\approx \sqrt{U/\omega ^{\prime }}$, on the bases of
(17) and (18), it is possible to present the condition (20) in the form $%
\sqrt{\omega ^{\prime }U}\approx \Omega $. Thus, in the case of overlapping
of resonances the phase oscillations frequency - $\omega _{ph}$, coincides
by an order of magnitude with the frequency distance between harmonics in
the pumping spectrum. Usually criterion of the dynamic stochasticity,
equivalent to the overlapping of resonances, is written down by introducing
the stochasticity coefficient 
\begin{equation}
K\approx \sqrt{\omega ^{\prime }U}/\Omega>1.  \label{21}
\end{equation}

In the range of statistical motion the nonlinear oscillating system is
described with the help of distribution function $\rho (t)$, for which it is
possible to obtain the diffusion equation \cite{1} 
\begin{equation}
\frac{\partial \rho }{\partial t}=D\frac{\partial ^{2}\rho (I,t)}{\partial
I^{2}},  \label{22}
\end{equation}
where $D=\frac{1}{2}U^{2}(I)T$ is a diffusion coefficient. Now from (22)
with the help of $\rho (I,t)$ it is easy to get $\overline{I}=\overline{I}%
_{0}+Dt$, where average is understood as a statistical average. Diffusion
growth of the action reduces the growth of $\sqrt{\langle x^{2}\rangle }$ in
the range of $x>x_{L}$. As an energy of the particle, located in a hole, is $%
E_{o}=I\omega _{o}$, then in the range of stochastic dynamics $\overline{E(t)%
}$=$E_{o}+\omega _{o}\sqrt{Dt}$ and ''heating'' of the particle takes the
place. The above-mentioned reasonings are proved by numerical calculations
(Fig. 5).

The first numerical experiments for stochastic "heating" of a nonlinear
oscillator were carried out long time ago, see \cite{1}.

The condition of resonances overlapping has visual interpretation on a phase
plane.

The mathematical pendulum, in an association with initial conditions, can
make two types of motion: oscillatory and rotary. They are separated by a
separatrix on the phase diagram. The overlapping of resonances on the phase
plane corresponds to a touch of separatrixes (Fig. 6), if the width of the
separatrix is estimated as $\omega^\prime\Delta I_n\approx\sqrt{%
\omega^\prime U}$ (in frequency units).

The dynamic stochasticity usually originates in a narrow layer near to a
separatrix \cite{1,2}. Therefore, at quantum reviewing we shall be
especially interested in quantum properties of the system near to the
separatrix. In other words, we shall be interested in a wave function of
isolated nonlinear resonance in the absence of overlapping. The analysis of
these properties can explain essence of a quantum chaos.

In conclusion of this paragraph we shall remark, that the condition of
overlapping of resonances depends on the action $\Delta I_n$ as the solution
of the equations generated by the universal Hamiltonian. Therefore, at
quantum reviewing, for establishment the criterion of overlapping of
resonances it is enough in (20) under $\Delta I_n$ to understand the
relevant magnitude obtained from the quantum equations generated by the
Hamiltonian (9).

Except the quantum estimation for $\Delta I$ we shall be interested also in
quantum dynamics near to the separatrix $(E\approx U)$, as the basis of
directed random motion in one-dimensional nonlinear systems.

\section{Quantum-mechanical consideration}

The quantummechanical consideration of isolated nonlinear resonance, as well
as overlapping of two resonances was presented in \cite{3,5,6}.In \cite{5}
it was specified, that the Schrodinger equation for a nonlinear isolated
resonance can be reduced to the Mathieu's equation, and if the condition of
the resonances overlapping is fulfilled the correlations drop \cite{6}
(numerical computational methods). We in this paragraph are interested
essentially in quantummechanical characteristics of the problem - wave
function and energy spectrum. We shall investigate, how can at quantum case
the unpredictability of hit of a system in any quantum state (analog of a
stochastic stratum near a separatrix in classical mechanics).

The universal Hamiltonian (9) depends on the basic parameter of nonlinear
oscillations, $\omega^\prime$. At quantum reviewing corresponding to the
universal Hamiltonian Schrodinger equation will be also depending on $%
\omega^\prime$. So we came to the quantum-mechanical consideration of the
nonlinear-oscillating system within the frame of approximation made in \S1.

The Schrodinger equation relevant to the Hamiltonian (9), is: 
\begin{equation}
\frac{d^2\Psi}{d\alpha^2}+\frac{2}{\chi}[E-U \cos\alpha]\Psi=0,  \label{23}
\end{equation}
where $\chi=\omega^\prime\hbar^2$.

Let us clarify essence of the parameter $\chi$. The value $\omega^\prime
\hbar$ is the quantum (the minimal portion) of the frequency shift
stipulated by nonlinearity (i.e., the frequency quantum of nonlinearity).
Hence, value $\chi=\omega^\prime \hbar^2$ is the energy quantum of
nonlinearity.

The equation (23) is the Mathieu's equation which analysis we shall make
below. For now we want to get quasi-classical wave functions relevant to an
approximation $\Lambda=U/\chi \gg 1$ of the Schrodinger equation. It is
known, that quasi-classical wave function is: 
\begin{equation}
\Psi (\alpha )=\frac{c}{\sqrt{\Delta I}}\exp \biggl(i/\hbar
\int\limits_{_{0}}^{\alpha }\Delta I(\alpha )d\alpha \biggr),  \label{24}
\end{equation}
where $c$ is the normalizing constant and $\Delta I$ can be find from the
integral of the energy 
\begin{equation}
\Delta I=\sqrt{2/\omega ^{\prime }(E+U\cos \alpha )}.  \label{25}
\end{equation}

Substituting (24) in (23), after integration we get: 
\begin{equation}  \label{26}
\hskip-0.5cm \Psi _{+}(\alpha )\!=\!1/2\frac{(E+U)^{1/4}}{\sqrt{K(k)}}\frac{%
\exp \biggl(2i\sqrt{(E+U)/\chi }E(\alpha /2,k)\biggr)}{(E+U\cos \alpha
)^{1/4}}, \\
\end{equation}
\[
E>U, 0\leq \alpha \leq \pi ; \newline
\Psi_{-}(\alpha )=1/\sqrt{2}\frac{(U/2)^{1/4}}{\sqrt{K(1/k)}}\times
\]
\begin{equation}  \label{27}
\times\frac{\exp \biggl(i\sqrt{\frac{2}{U\chi}}[(E-U)F(\gamma,1/k)+2UE(%
\gamma,1/k)]\biggr)} {(E+U\cos \alpha )^{1/4}}, \\
\end{equation}
\[
0<E<U,\;\;\;0\leq \alpha \leq \alpha _{o};
\]
\begin{equation}
\gamma=\arcsin \sqrt{\frac{U(1-\cos \alpha )}{U+E}},\alpha _{o}=\arccos
(-E/U),  \label{28}
\end{equation}
where $F(\ldots )$ is the elliptic \ integral of the first kind,$E(\ldots )$
is the elliptic integral of the second kind and $K(\ldots )$ is the first
kind complete elliptic integral.

Let's note, that as in the considered case $\Lambda \gg 1$ quasi-classical
wave functions (26) and (27) oscillate fast with the variation of $\alpha$,
having peaks at turning points $\pm\alpha_o$. It is a common property of
wave functions in a quasi-classical approximation.

The wave functions corresponding to the separatrix can be obtained in the
limit $E\rightarrow U$: 
\[
\Psi_-(\alpha)=\Psi_+(\alpha)=\Psi_s(\alpha)=
\]
\begin{equation}
=1/2\sqrt{2}\frac{1} {\ln4\sqrt{\frac{U}{|E-U|}}} \frac{\exp(i2\sqrt{2\Lambda%
}\sin\alpha/2)}{(1+\cos\alpha)^{1/4}}.  \label{29}
\end{equation}

According to expression (29) for a wave function near the separatrix the
frequiency of fast oscillations practically does not vary. The turning
points $\alpha_o\langle \pi$ approach to $\pm \pi$ and the peaks of fast
oscillations become negligible low. This is connected with the
logaritmically diverging factor in (29). With the help of (29)it is easy to
find an equation for nodal points of separatrix wave function, $%
\Psi_S(\alpha)=0$ 
\begin{equation}
2\sqrt{2U/\chi}\sin\frac{\alpha_n}{2}=\pi/2+2\pi n, n=1,2,\ldots  \label{30}
\end{equation}

Differentiating the relation (24) it is possible to calculate the density of
nodal points 
\begin{equation}
\frac{dn}{d\alpha_n}=\frac{1}{2\pi}\sqrt{\frac{2U} {\chi}}\cos\frac{\alpha_n%
}{2}.  \label{31}
\end{equation}

The high density of nodal points is provided by the major parameter of
quasi-classical consideration $(2U/\chi )^{1/2}$,which is suppressed by the
zeroes of the factor $\cos \frac{\alpha _{n}}{2}$ in the points $\alpha
_{n}=\pm \pi $ (see Fig.7). Another important characteristic of the quantum
state near to the separatrix is the density of energy levels. Bohr -
Zommerfeld quantization condition looks like: 
\begin{equation}
I=\oint \Delta I(\alpha )d\alpha =n\hbar .  \label{32}
\end{equation}

Taking into account (25), we get: 
\[
I=\sqrt{2/\omega^\prime}\int\sqrt{E+U\cos\alpha}d\alpha=
\]
\begin{equation}
=\frac{2}{\sqrt{U\chi}}\biggl((E-U)K(1/k)+2U E(1/k)\biggr)=n.  \label{33}
\end{equation}

With the help of (33) for the density of energy levels it is possible to
obtain: 
\begin{equation}
\frac{dn}{dE}=\frac {1}{\sqrt{U\chi}}K(1/k)  \label{34}
\end{equation}

In the limit $E\rightarrow U$ for the energy level density, $dn/dE$, we get: 
\begin{equation}
\biggl(\frac{dn}{dE}\biggr)_{s}=\frac{1}{\sqrt{\chi U}}\ln4 \sqrt{\frac{2U}{%
|E-U|}}.  \label{35}
\end{equation}

As it can be seen from (35) the level densities are logarithmic divergent
near the separatrix.

Now let us try to evaluate quasi-classical condition of the overlapping of
resonances. For this purpose it is necessary to calculate average values $%
\Delta I\ $ for a halfcycle of motion both above and lower of the
separatrixes. Taking into account (25)-(27), we obtain: 
\begin{equation}  \label{36}
\langle \Delta I \rangle ^{+}=\int\limits_{-\pi }^{\pi }|\Psi
_{+}(\alpha)|^{2}\Delta I(\alpha )d\alpha =\pi /2\sqrt{\frac{E+U}{\chi}} 
\frac{\hbar}{K(k)} \\
\end{equation}

\begin{equation}
\langle \Delta I \rangle^{-}=\int\limits_{-\alpha _{o}}^{\alpha _{o}}|\Psi
_{-}(\alpha )|^{2}\Delta I(\alpha )d\alpha =\sqrt{\frac{U}{2\chi}}\frac{%
\alpha _{o}\hbar}{K(1/k)}  \label{37}
\end{equation}
and near the separatrixes, $E\approx U$: 
\begin{equation}
{\langle \Delta I\rangle }^{+}\approx {\ \langle \Delta I\rangle }^{-}\approx%
{\langle \Delta I\rangle }_{S}= \pi \sqrt{U/2\chi }\frac{\hbar} {\ln 4\sqrt{%
\frac{2U}{|E-U|}}},  \label{38}
\end{equation}
where $\langle \ldots \rangle ^{\pm }$ denotes averaging which the help of
wave functions (26), (27) and (29) . Let's note, that 
\[
\langle\Delta I\rangle^-/\hbar=(U/2\chi)^{1/2}(\alpha_{o}/K(1/k))\approx
\Lambda^{1/2} \gg 1
\]
is in a good agreement with a quasi-classical condition of motion. Because
of the coincidence of classical values $\overline{ \Delta I}_{\pm}$ with the
quasi-classical ones $\langle \Delta I \rangle^{\pm}$ it is natural that
conditions of overlapping resonances will also coincide $\omega^\prime 
\overline {\Delta I}_-=\omega^\prime\langle{\Delta I}\rangle^-=\Omega $.
Hence, one can conclude that under quasi-classical conditions $\Lambda\gg1$
and the condition of overlapping the resonances the stochastic "heating" of
electron being under the conditions given in section 1 and the obtaining of
high excitations can be carried out(see classical case,Fig.5).
Quasi-classical expressions (36),(37) by the form coincide with the similar
classical expressions (17)and (18). But as opposite to the classical
formulas, in quasi-classical expressions under $E$ should be understood $E_n$
energy spectrum of quasi-classical levels. With the help(33)and (34) it is
possible to determine number of the levels entrapped in a nonlinear
resonance: 
\begin{equation}
\Delta n =\frac{dn}{dE}\Delta E,  \label{39}
\end{equation}
where 
\begin{equation}
\Delta E=\hbar \omega^{\prime}\langle\Delta I\rangle^-.  \label{40}
\end{equation}
$\Delta n$ is the important characteristics of an isolated nonlinear
resonance. Using (34), (37),(39) and (40)it is easy to show, that the number
of levels entrapped in nonlinear resonance is $\Delta n \simeq \frac{\alpha_o%
}{2}$. According to (35), (38),(39) and (40) it is also easy to show, that
near the separatrix there is $\Delta n_s\simeq\frac{\pi}{2}$. While
according to (34) and (35) the density of levels increases logarithmically
with approaching to the separatrix, the number of levels $\Delta n$
entrapped in resonance is not increased. This is caused by the sharp fall of
the action variation near $U \simeq E$ (see Fig.4). Thus the number of
entrapped levels in a nonlinear resonance is not great and remains the same
when approaching to the separatrix. Major values $\Delta n$, as it was shown
in \cite{6} by means of numerical methods, can be reached in case of
overlapping resonances.

Let's analyse the Schodinger equation (23). We assume, that $U$ and $\chi$
are the values of one order and that is why the quasi-classical
approximation can not be used.

In the limit case $E\gg U$ it is possible to use limiting $(U\rightarrow 0)$
formulas for the Mathieu's functions \cite{9}: 
\[
\Psi_0(\alpha)=ce_o(\alpha)=1/\sqrt{2},
\]
\[
\Psi_n(\alpha)=ce_n(\alpha)=\cos n\alpha,
\]
\begin{equation}
\Psi_n(\alpha)=se_n(\alpha)=\sin n\alpha,n=1,2,3\ldots.  \label{41}
\end{equation}

These functions should satisfy equation 
\begin{equation}
\frac{d^{2}\Psi }{d\alpha ^{2}}+\frac{2E}{\chi }\Psi =0  \label{42}
\end{equation}
which follows from (23) at $U\rightarrow 0$. The equation describes harmonic
oscillations with the frequency $\sqrt{2E/\chi }$. To reduce it in the
correspondence with the solution (41) it is necessary to require 
\begin{equation}
\sqrt{\frac{2E}{\chi }}=n\;\;\;\text{or}\;\;\;E_{n}=1/2\chi n^{2},
\label{43}
\end{equation}
where $n=1,2,3\ldots $. The last relation leads to the energy spectrum
quadratically depending on the quantum number.

It is possible to use the relations obtained with the help of an averaged
universal Hamiltonian for calculation of $\Delta I$ in the zero order with
respect to the $U/E$ 
\begin{equation}
\langle(\Delta I)^2\rangle_n=\frac{2}{\omega^\prime}E_n.  \label{44}
\end{equation}

Then taking into account of (43) we get 
\begin{equation}
\langle\Delta I\rangle_n\approx\sqrt{\langle(\Delta I)^2\rangle_n} =n\hbar.
\label{45}
\end{equation}

It is possible to obtain the condition of the overlapping of resonances for
a frequency shift, caused by the variation of an action 
\begin{equation}
\delta\omega\approx\omega^\prime\langle\Delta I\rangle_n=\omega^\prime
n\hbar\geq \Omega.  \label{46}
\end{equation}

Using (46) the condition of the resonances overlapping could be written as $%
\delta \omega \approx\omega^{\prime}n\hbar \geq\Omega $. It is significantly
difficult to fulfill this condition as compared with similar one of
quasi-classical case, because it requires excessively small $\Omega$.

In the opposite limit case $E \ll U$ we have condition of overlapping the
resonances, which is also difficult to be fulfilled and by this reason we
don't present a detailed analysis.

It is well known, that eigenvalues of the Mathieu's equation can be defined
by means of the Ince-Strutt diagrams (Fig.9) constructed for the first time
to studying a parametrical resonance \cite{9}. As follows from these
diagrams, each value $U$ corresponds to set of eigenvalues $E_n$ and
periodical wave functions $ce_n(\alpha),se_n(\alpha)$. Curvs in Fig.9
corresponding to the realized quantum states are known
Mathieu-characteristics. Essential feature of Mathieu-characteristics is the
presence of points of branching in the neighbourhood of a line $U=E$,
corresponding to the classical separatrix. Moving along
Mathieu-characteristics from left to right in points of branching being at
the left of separatrix line disappears two-fold degeneration. So through the
passing of separatrix line $U=E$ and reaching the point of branching, to the
right of a separatrix wave functions merge again but now - $se_n$ and $%
ce_{n-1}$. The emergence of such picture (point of branching from two sides
of the separatrix) at the passage to quantum consideration is a principal
characteristic describing a quantum system near classical separatrix. One
can observe the appearance of unpredictability of occupied quantum levels
with the help of branching points located on both sides of a separatrix. Let
us suppose, that one of the system parameters, for example, the amplitude of
variable field $U$ varies adiabatically ($U\rightarrow U+\epsilon cos \kappa
t,\epsilon \ll 1$, $\kappa$-frequency of slow motion). In common problems of
a quantum mechanics is believed that the distance between levels in an
energy distribution, depending on exterior parameters vary synchronously
with adiabatically varying parameter, not changing in this case a quantum
state. The situation is changed radically, if in a quantum mechanical
problem of the definition of energy distribution and eigenfunctions is
reduced to the analysis of diagrams of the Ince-Strutt type. At slow moving
along the curves of Mathieu-characteristics due to the adiabatic change of
the amplitude of variable field, after multiple passages through the
branching points it is impossible to determine exactly in which
Mathieu-characteristic (in which quantum state) the system can be found. One
can consider this appearance as quantum analog of formation of a stochastic
layer of motion in the area of classical separatrix.

Let us pass to the analysis of overlapping criterion in a quantum
case.Taking into account the quantum virial theorem, connecting average
kinetic energy $T$ with the average potential energy $U$ 
\begin{equation}
2<T>=<\alpha \frac{dU}{d\alpha }>,  \label{47}
\end{equation}
for variation of action $\Delta I$ with the help of (44) we obtain: 
\[
<\Delta I>\approx \sqrt{(\Delta I)^2)}=\lambda (\Lambda)\sqrt{\omega
^{\prime }U}
\]
where 
\begin{equation}
\lambda(\Lambda)=\sqrt{-\int_{0}^{\pi }ce_{n}^{2}(\alpha ,\Lambda)\alpha
\sin \alpha d\alpha }\text{ \ , \ }  \label{48}
\end{equation}
and $\Lambda=U/\chi $. The condition of overlapping of resonances takes the
form:

\begin{equation}
\delta \omega =\omega ^{\prime }<\Delta I>\approx \lambda(\Lambda)\sqrt{%
\omega ^{\prime }U}\geq \Omega \text{ , }  \label{49}
\end{equation}
or 
\begin{equation}
\lambda(\Lambda) K\geq 1\text{ , }  \label{50}
\end{equation}

Comparing the quantum criterion (50) with the classical (21), one can
conclude that in quantum case there is additional quantum factor $\lambda
(\Lambda)$.

As can be seen from (50), conditions of resonances overlapping in the
quantum case is determined by magnitude of $\Lambda$ .This condition is
reduced to that how many levels of nonlinearity of energy $\chi$ can be
located in $U$.

Thus, at quantum reviewing, the additional factor $\lambda(\Lambda)$ appears
in the overlaping criterion. The physical sense of $\lambda(\Lambda)$ can be
clarified. If the classic criteria of the resonances overlapping $\sqrt{%
\omega^\prime U} \approx\Omega$, is fulfilled then, according to (50), the
magnitude of $\lambda(\Lambda)$ determines the conditions of weakening $%
(\lambda(\Lambda)<1)$ or amplifying $(\lambda(\Lambda)>1)$ the resonances
overlapping criterion, in the comparison with the classical one. Differently 
$\lambda(\Lambda) >1$ means that the condition of overlapping of resonances
more easily can be fulfilled in a quantum case.

The data of numerical calculations for the magnitude $\lambda(\Lambda)$ are
given on the Fig 10. As it is visible from Fig. 10, $\lambda(\Lambda) >1$,
i.e., the resonances overlapping criterion amplifies in the quantum case.
For high values $\Lambda$ (the quasi-classical case) $\lambda(\Lambda)$
should tend to unit. On Fig. 11 the eigenvalues of an energy $E/\chi$ as a
functions of $U_o$ are presented for states described by the functions $%
ce_4(\alpha,U)$ and $ce_6(\alpha,U)$. From Figs. 10 and 11 it is easy to see
that the maximum value of $\lambda(\Lambda)$ corresponds to $\Big(\frac{%
\partial E}{\partial U}\Big)_n=0$. Thus, in the quantum case the area, lying
below classical separatrix, corresponds to the area of the maximum
stochasticity.

%
%
\begin{figure}[tbp]
\caption{Continuous line corresponds to the anharmonic potential, $%
U_{NL}(x)= \frac {m\protect\omega_o^2x^2}{2}+\protect\gamma x^3+\protect\beta
x^4$. The dashed line corresponds to the harmonic potential $U_L(x)=\frac{m%
\protect\omega_o^2x^2}{2}$.They coincide up to point $x_L$}
\label{fig1}
\end{figure}

\begin{figure}[tbp]
\caption{Periodic series of rectangular pulses. $\protect\tau$ is the pulse
length, $T$ is the recurring period. }
\label{fig2}
\end{figure}

\begin{figure}[tbp]
\caption{Pumping spectrum $f(t)$ consists from many harmonics, multiples to
the $\Omega= 2\protect\pi/T$, enveloping frequency range from $\protect\omega%
_o+1/\protect\tau$ up to $\protect\omega_o1/\protect\tau$. }
\label{fig3}
\end{figure}

\begin{figure}[tbp]
\caption{$\overline{\Delta I_-}$ as a function of ratio $E/U$ in the
classical case.}
\label{fig4}
\end{figure}

\begin{figure}[tbp]
\caption{Diffusion growth of the action obtained at the values of parameters 
$f_o=0.5, \protect\omega_o=20,x_l=1, \Omega=0.2,T=10, \protect\tau=1$ and
the stochasticity factor $K=\frac{\protect\tau}{T}f_o\protect\omega%
^{\prime}T\approx3$. }
\label{fig5}
\end{figure}

\begin{figure}[tbp]
\caption{Phase trajectories of the mathematical pendulum near to two
resonances .}
\label{fig6}
\end{figure}

\begin{figure}[tbp]
\caption{Quasiclassical wave function below the separatrix. $E=0.9 U,\Lambda
\sim 100$}
\label{fig7}
\end{figure}

\begin{figure}[tbp]
\caption{Quasiclassical wave function near the separatrix. $E\sim U.$}
\label{fig8}
\end{figure}

\begin{figure}[tbp]
\caption{Dependence of the eigenvalue $E$ from the $U$ for different
Mathieu's functions. Dashed line corresponds to the separatrix. Symbol $\circ
$ denotes branching points above the separatrix,symbol $\bullet$ - branching
points below the separatrix }
\label{fig9}
\end{figure}

\begin{figure}[tbp]
\caption{Parameter $\protect\lambda$ as a function of $\Lambda$ for
different Mathieu's functions $ce_4 ce_6$.}
\label{fig10}
\end{figure}

\begin{figure}[tbp]
\caption{Eigenvalue $E$ as functions of $U_o$ for Mathieu's functions $ce_4
ce_6$. Dashed line corresponds to the separatrix. }
\label{fig11}
\end{figure}

%
%


\begin{references}
\bibitem{1}  R. Z. Sagdeev, D.A.Usikov, G. M. Zaslavsky, Nonlinear Physics.
New-York: Hardwood Acad.Publ., 1988.

\bibitem{2}  A. J. Lichtenberg, M. A. Lieberman, Regular and Stochastic
Motion, Springer-Verlag, New-York, Heidelberg, Berlin, 1983.

\bibitem{3}  G. M. Zaslavsky, Phys. Reports v.80,p.157,1981.

\bibitem{4}  F.Haake,Quantum Signatures of Chaos,Springer,Berlin 2001.

\bibitem{5}  G.P.Berman, G. M. Zaslavsky, Phys. Lett.,v.61A, p295, 1977.

\bibitem{6}  G.P.Berman, G. M. Zaslavsky and A.R.Kolovsky, Phys.
Lett.,v.87A, p152,1982.

\bibitem{7}  M. C. Gutzwiller, Chaos in Classical and Quantum Mechanics,
Springer, New York, 1990.

\bibitem{8}  K. T. Alligood, T. D. Sauer and J. A. York, Chaos an
introduction to Dynamical Systems Springer, New York 1996.

\bibitem{9}  Handbook of Mathematical Functions Edited by M.Abramovitz,
National Bureau of standards Appl.Math. series.55 , Issued June 1964.
\end{references}
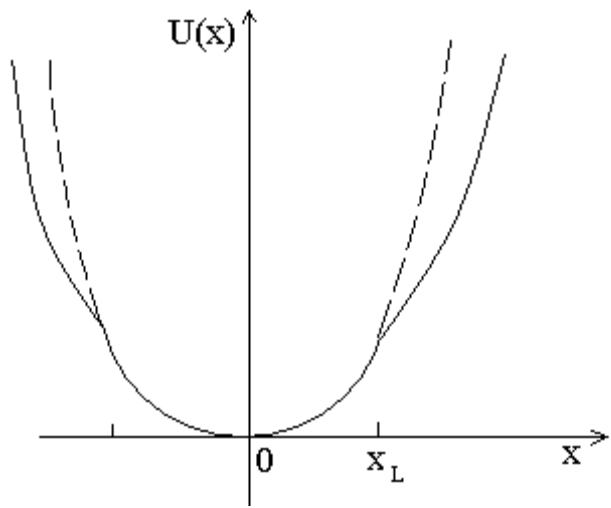
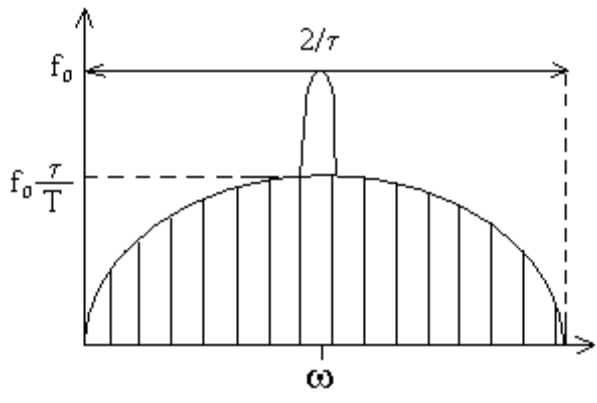
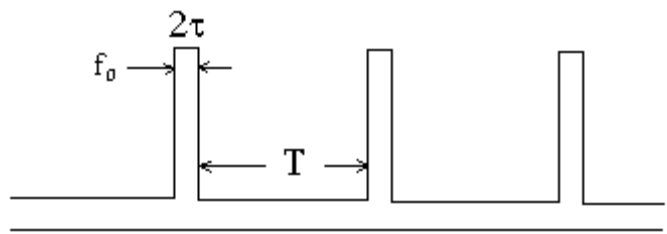
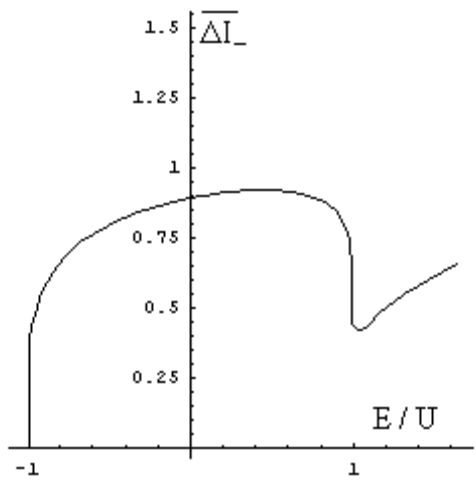
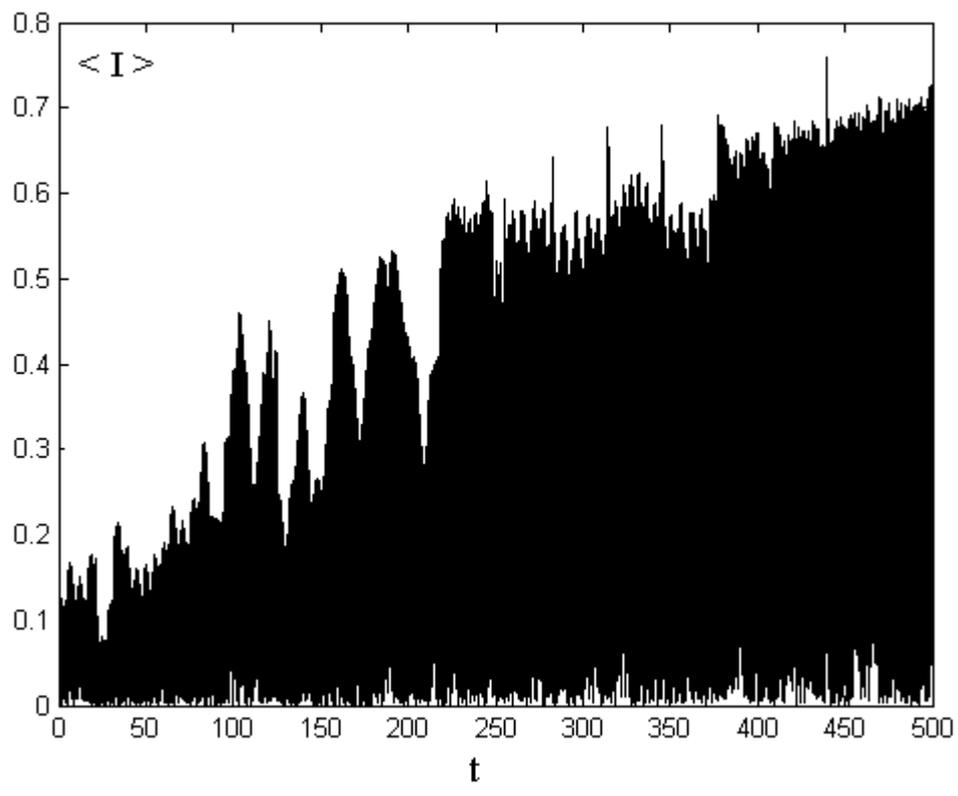
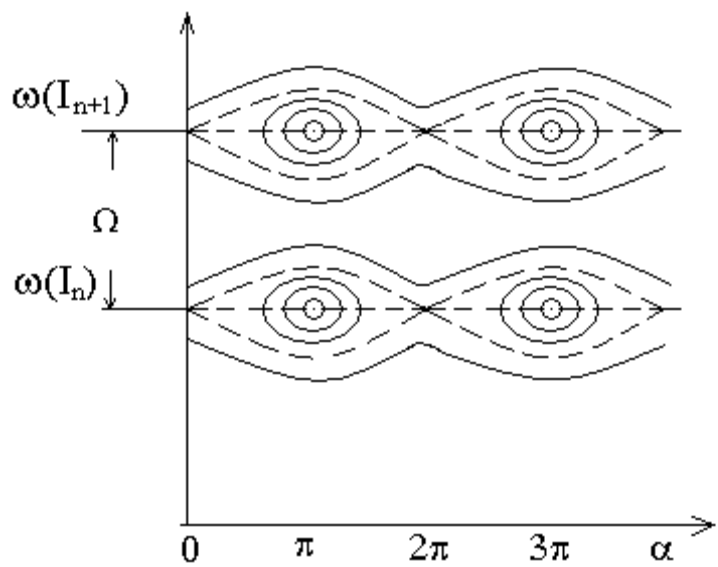
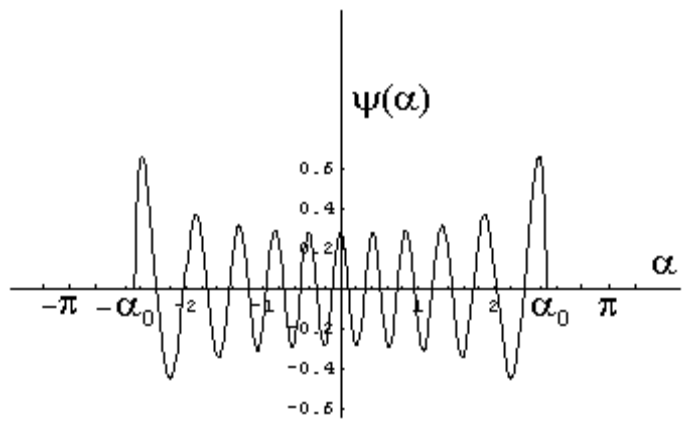
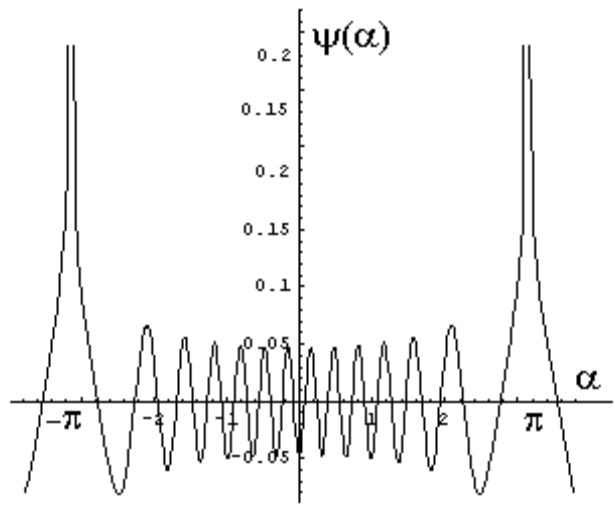
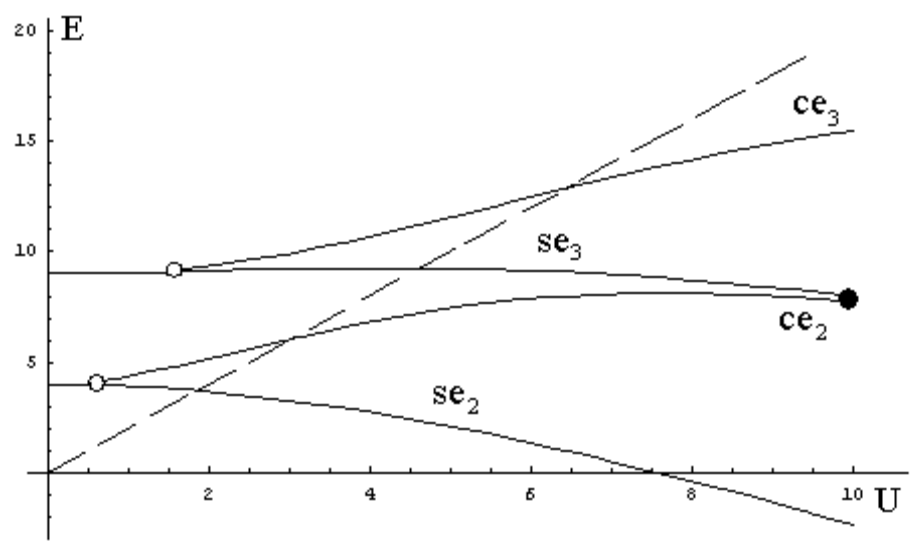
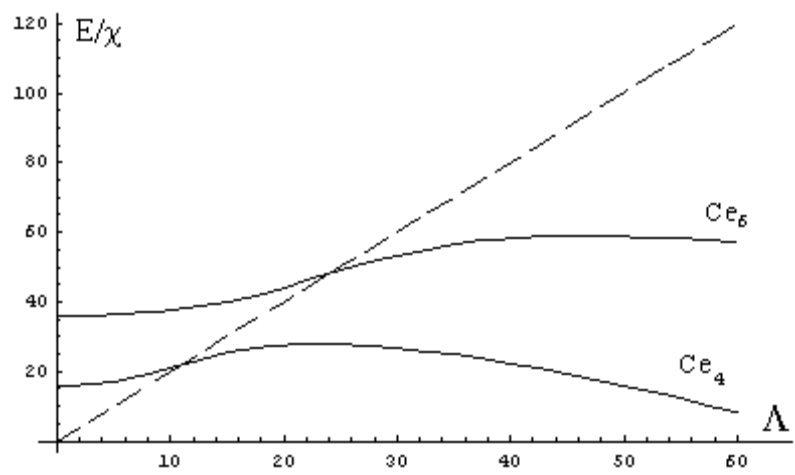
\end{document}